\documentstyle[twoside,fleqn,espcrc2,epsf]{article}

\newcommand{\AmS}{{\protect\the\textfont2
  A\kern-.1667em\lower.5ex\hbox{M}\kern-.125emS}}

\def\lesssim{\lower.7ex\hbox{${\buildrel < \over \sim}$}}
\def\gtrsim{\lower.7ex\hbox{${\buildrel > \over \sim}$}}

\title{Uncertainty of the atmospheric neutrino fluxes}
\author{M.Honda\address{Institute for Cosmic Ray Research, 
University of Tokyo,\\ Tanashi 3-2-1,  Tokyo 188 JAPAN}
}
\begin{document}
\begin{abstract}
The uncertainty in the calculation of atmospheric neutrino fluxes is 
studied.
The absolute value of atmospheric neutrino fluxes is sensitive to 
variation of the primary cosmic ray flux model and/or the interaction 
model.
However, 
the ratios between different kind of neutrinos stay almost unchanged with 
these variations.
It is unlikely that the anomalous ratio 
$(\nu_{\mu}/\nu_e)_{obs}/(\nu_{\mu}/\nu_e)_{MC}$
reported by Kamiokande and Super Kamiokande is caused by 
the uncertainty of predicted atmospheric neutrino fluxes.
\end{abstract}
\date{}
\vskip 2mm

\maketitle

\section{Introduction}

After the discovery of an anomalous ratio of $\nu_e/\nu_\mu$
in the atmospheric neutrinos by Kamiokande,
refined calculations of atmospheric neutrino fluxes have been
made intensively by several 
authors\cite{bn}\cite{bgs}\cite{hkhm}\cite{LK}\cite{HKKM}.
The differences between different authors have been compared 
in Ref.~\cite{compflux}.
The major differences are in the primary cosmic ray flux and 
in the hadronic interaction model.
These are also the main sources of uncertainty in the calculation
of atmospheric neutrino fluxes.

There have been many measurements of the primary cosmic ray flux, 
but they  do not agree with each other. 
We have to consider this disagreement as the uncertainty of the 
primary cosmic ray flux.
There also have been many experimental studies of hadronic interactions.
However, the most important piece of information, 
the energy spectrum of secondary particles in the projectile region,
is not well known enough for a accurate calculation of atmospheric neutrino
fluxes.
This is partly because, as high energy physics experiments 
moved to the colliders,
this measurement has become more difficult to make.

Adding to the above, the density structure of the atmosphere is a 
potential source of uncertainty. 
In many calculations, the US standard is often used as the `standard'
model.
However, it is also known that the atmospheric density structure 
has a latitude dependence and seasonal variations.

In this paper, we study how these uncertainties affect the 
atmospheric neutrino calculation in the energy range of 
\lesssim 10~GeV based on Ref.~\cite{HKKM}(HKKM).
Note we use the one-dimensional approximation as HKKM throughout this
paper.

\begin{figure}[ht]
\centerline{\epsffile{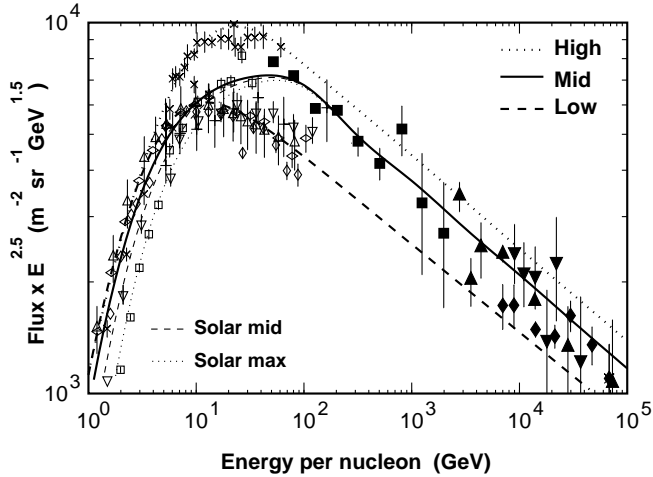}}
\caption{
{
Observed cosmic ray proton flux.
Crosses from Ref.~\cite{webber79},
open upward triangles stand for LEAP\cite{leap}, 
open squares for MASS\cite{mass},
open downward triangles for IMAX\cite{imax},
open vertical diamonds for CAPRICE\cite{caprice}, and
open horizontal diamonds for BESS\cite{bess}.
pluses,
closed squares, 
closed vertical diamond,
closed upward triangles, and
closed downward triangles
are from Refs[22],[26],[25],[34],[35],[36], and [38] of HKKM 
respectively.
}
}
\label{le10000}
\end{figure}

\section{Variation of calculation Model}
\subsection{Primary cosmic ray flux}
\label{sec-1ry-vari}

In Fig.~\ref{le10000} are shown the observations of cosmic 
ray protons at $< 10^4$~GeV by different groups.
We consider 3 flux models in this study: high, mid, and low
shown in the figure for solar min. 
The flux models for primary cosmic rays agree in $\lesssim 5$~GeV region, 
where the differences are small except for these caused by solar 
modulation.

Cosmic rays above a few 100 MeV originate in Galactic space. 
When they enter the solar sphere, they are pushed back by the
solar wind.
As this effect is more pronounced for lower energy cosmic rays,
the energy spectrum of low energy cosmic rays varies 
with the strength of the solar wind, or with the solar activity.
However, this modulation is expected to be around 5\% from the minimum to 
the maximum of solar activity at 10 GeV. 
Above this energy, the effect of solar activity on the 
cosmic ray flux is very small.

The variation due to solar activity in the $\lesssim 10$~GeV region
agrees well with the expectation based on the Mid-flux model and
conventional solar modulation formulae.
The exception in Fig.~\ref{le10000} is the crosses 
which are the base data for the high flux model.
We consider that the uncertainty in primary cosmic ray flux is 
small for $\lesssim 10$~GeV.

Above 10 GeV, there are rather large differences among the different 
experiments. 
This may be caused by the inherent difficulties in making the measurements.
Particularly, the problems in the estimation of instrumental 
efficiency and limited exposure time in balloon experiments.
At this moment we have to assume that there are large uncertainties
in the primary cosmic ray flux measurements in the $\gtrsim $10 GeV region.
However, it is interesting that recent experimental values
with super conductive magnetic spectrometers distribute rather 
around the Low flux model.

It is noted that around 20\% of nucleons are carried by nuclei
heavier than the proton.
We do not show the details but for the heavier nucleus
cosmic rays we use the flux value of HKKM.

Even using the one dimensional approximation, 
the geomagnetic field affects the calculation of atmospheric neutrino 
fluxes through the primary comsic ray flux due to the geomagnetic cutoff.
Principally, however, the geomagnetic cutoff is not a source of uncertainty 
since it can be calculated to very good accuracy from the measured 
geomagnetic field.

\subsection{Hadronic interaction}
\label{sec-int-vari}

In the calculation of the atmospheric neutrino fluxes,
hadronic interaction Monte Carlo code is one of the most important 
components.
However, there are uncertainties in the hadronic interaction model,
due to the lack of suitable experimental data.
The interaction models used by different calculations are
slightly different to each other.

\begin{figure}[ht]
\centerline{\epsffile{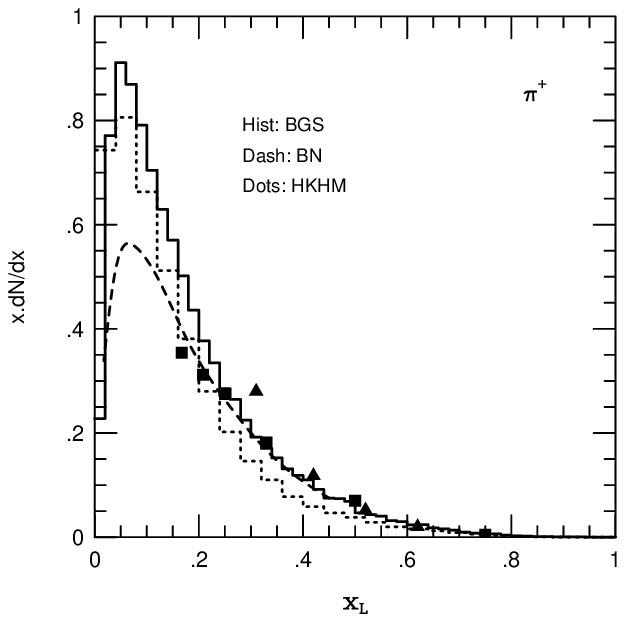}}
\caption{Secondary $\pi^+$ energy spectrum calculated using 
the interaction model
used by various different authors
(taken from  Ref. \cite{compflux}).}
\label{stanev1}
\end{figure}

\begin{figure}[ht]
\centerline{\epsffile{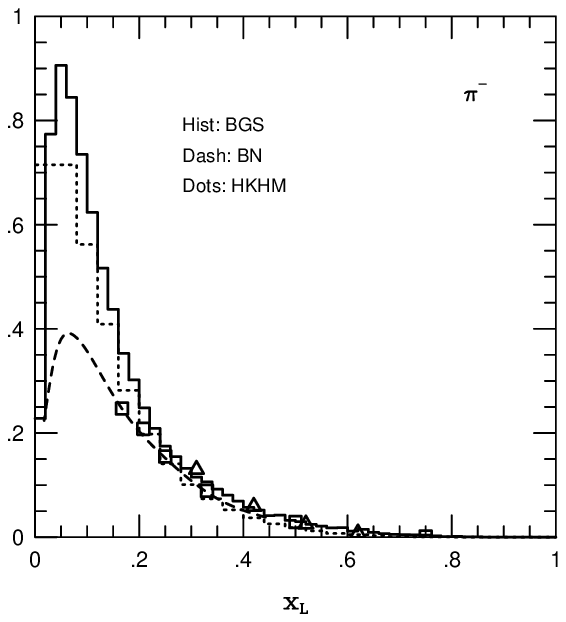}}
\caption{Secondary $\pi^-$ energy spectrum calculated using 
the interaction model 
used by various different authors
(taken from  Ref. \cite{compflux}).}
\label{stanev2}
\end{figure}

A comparison of the secondary particle momentum spectrum is shown in 
Fig.~\ref{stanev1} and in Fig.~\ref{stanev2} 
for several Monte Carlo codes which have been 
used in the calculation of atmospheric neutrino fluxes. 
(This figure is taken from Ref. \cite{compflux}).
The variable $x$ used in the figure is defined as 
$x = P_{2ndary}/P_{incident}$.
It should be noted that the difference between the histograms
(BGS\cite{bgs} and HKHM\cite{hkhm}) and the smooth curve (BN\cite{bn}) is 
not so large as it appears.
Since each value is multiplied by the variable $x$, 
their differences are emphasized in the low $x$ region.

It is generally difficult to modify an established hadronic interaction 
Monte Carlo code due to the energy momentum conservation and to the 
correlation of secondary particles.
However, in the case of the atmospheric neutrinos, the correlations are not 
important.
We use an `inclusive' hadronic interaction Monte Carlo code for the 
study of the variation of hadronic interactions.

The calculation of atmospheric neutrino fluxes with the inclusive 
interaction code is similar to the analytic calculation.
In the case of the analytic calculation, however, it is difficult to treat the
competition process between decay and interaction for hadronic particles.
This is the most crucial point in the analytic calculation.
Therefore, we make the Monte Carlo study with the inclusive 
hadronic interaction code in this paper.

Starting with the hadronic interaction model of FRITIOF 1.6 and Jetset 6.3, 
we consider `moderate' variations from that.
The variation of secondary particle energy spectrum is created as
follows: denoting a `starting' energy spectrum of secondary particles as
$$
{dN \over dx} = f_{ptl}(x)\ ,
$$ 
\noindent
where we take $x = {\rm E}_{2nd}/{\rm E}_{inc}$ and E is the kinetic 
energy.
Moderate variation to this secondary particle 
energy spectrum can be made with a parameter $\alpha$ as
$$
{dN \over dx} = A(\alpha)f_{ptl}(x^{1+\alpha})\ ,
$$
\noindent
where $A(\alpha)$ is a factor introduced for energy conservation as:
$$
A(\alpha) = {\int x f_{ptl}(x) dx \over \int x f_{ptl}(x^{1+\alpha}) dx } \ .
$$
\noindent
In other words, 
we request that the energy sum for each kind of particle is conserved. 
This modification method works efficiently to vary the slope of the 
secondary particle spectrum at $x\approx 0.3$.

\begin{figure}[ht]
\centerline{\epsffile{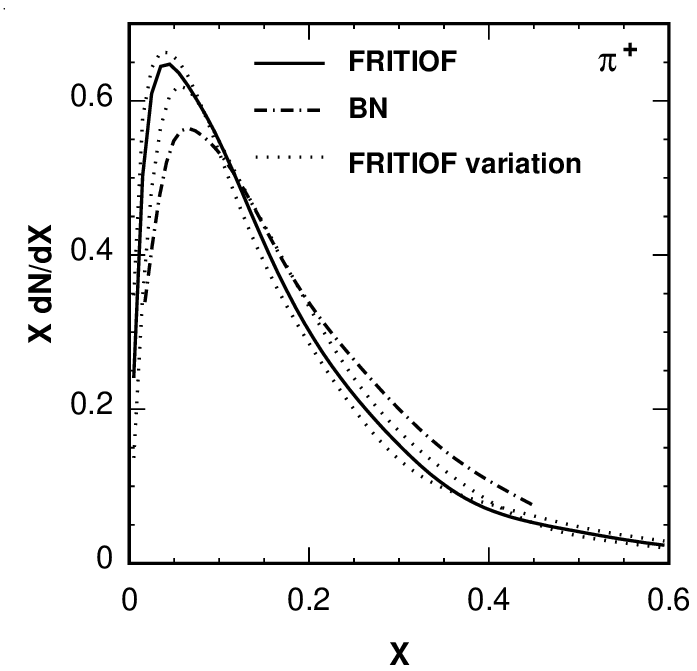}}
\caption{
Secondary $\pi^+$ energy spectrum from FRITIOF 1.6 and Jetset 6.3 
at 30~GeV 
with its variations for $\alpha=\pm 0.2$. For comparison, 
that of BN at $\approx 20$~GeV is also shown.
}
\label{test-pi+}
\end{figure}

\begin{figure}[ht]
\centerline{\epsffile{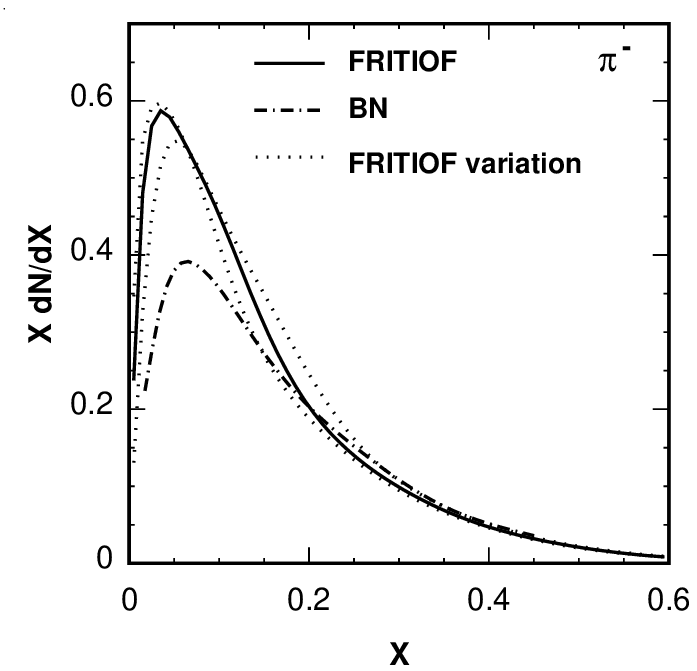}}
\caption{
Secondary $\pi^-$ energy spectrum from FRITIOF 1.6 and Jetset 6.3 
at 30~GeV,
with it's variations for $\alpha=\pm 0.2$. For comparison, 
that of BN at $\approx 20$~GeV is also shown.
}
\label{test-pi-}
\end{figure}

This modification also changes the number of particles for each
kind of secondary particle ($=\int x f_{ptl}(x) dx$).
The variation of $\alpha=\pm 0.1$ roughly corresponds to 
$\pm 10$~\% variation at $x=0.3$, and to a $\pm 20$~\% variation of 
the multiplicity. The variation of $\alpha$ from $-0.2$ to $+0.2$
is considered in this paper. The variation limit of $\alpha$ is
set as 2 times of the possible uncertainty of 
multiplicity ($\pm 10$~\%) in the hadronic interaction.

Examples of this modification are shown in 
Fig.~\ref{test-pi+} and Fig.~\ref{test-pi-} 
for $p + air \rightarrow \pi^\pm$ at $E_p = 32$~GeV
with the secondary particle spectrum of BN 
at $\approx 20$~GeV from Ref.~\cite{compflux} for comparison. 
One may notice that the secondary particle spectrum of BN is outside 
of our variation limit, even considering the difference of 
incident particle energy.
To include BN secondary energy spectrum, probably a 
different starting point is needed.

In addition, variations of inelastic cross-sections 
from $-20$~\% to $+20$~\% and $k/\pi$-ratio from 
$-40$~\% to $+40$~\% are also considered.
Note that we have assumed the same $\alpha$ for all kinds of secondary 
particles, although $f_{ptl}(x)$ is different for each particle, and
$\sigma_{inel}$ varies at the same ratio for all hadronic interacting
particles.
The variation range considered here is larger than the 
uncertainty normally considered\cite{pdg}.

\subsection{Density structure of the atmosphere}
\label{sec-atm-vari}

For the density structure of the atmosphere, 
the US-standard atmosphere model is often used.
However, since we are interested in the effect of the 
variation of atmospheric structure, we
introduce a simple model with a single scale height.
The air density is expressed as a function of height as:
$$
\rho = {\rho_{col}\over h_0} \exp(-{h \over h_0})\ .
$$
For the `standard', we take $\rho_{col}=1.231~{\rm kg/m^3}$ and 
$h_0 = 8.4$~km, such that it agrees with the US-standard in global
features.

We study the effect of variations from $-10$~\% to $+10$~\% both 
in the scale height and in the column density.
However, when we study the variation of interaction model only,
we use the US-standard atmosphere model.

\section{Variation of neutrino fluxes}

\begin{figure}[ht]
\centerline{\epsffile{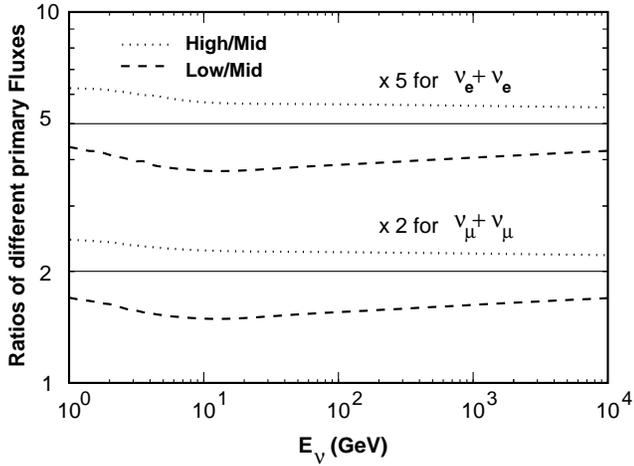}}
\caption{
The variation of atmospheric neutrino fluxes 
due to the change of primary cosmic ray model.
The ratio to that with Mid primary flux model (HKKM) is shown.
For the flux models, see the text in section \ref{sec-1ry-vari}.}
\label{1ryvari}
\end{figure}

\begin{figure}[ht]
\centerline{\epsffile{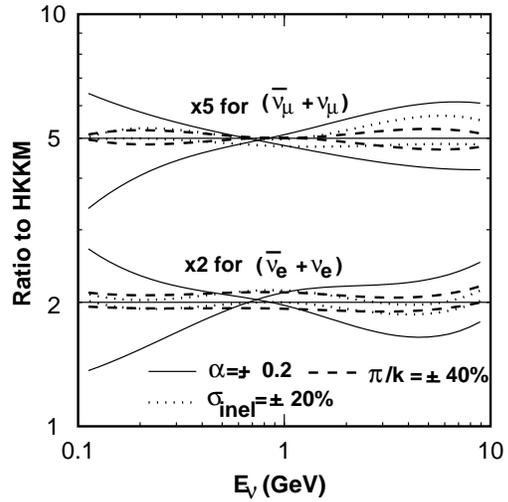}}
\caption{The variation of atmospheric neutrino fluxes due to the
variation of interaction model.
The ratio to the HKKM neutrino fluxes is shown.
For the parameters used in the figure, see the text in
section \ref{sec-int-vari}.}
\label{int2vari}
\end{figure}

\begin{figure}[ht]
\centerline{\epsffile{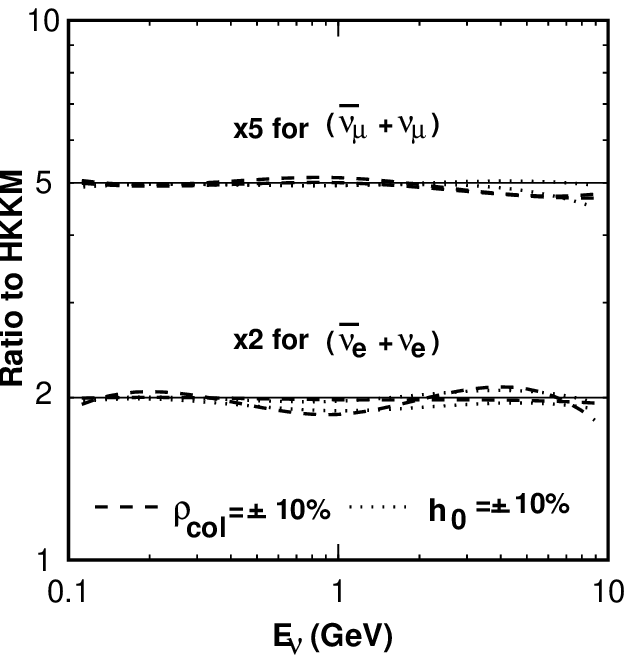}}
\caption{The variations of atmospheric neutrino fluxes due to the
variation of atmospheric model.
The ratio to the `standard' model is shown.
For the parameters used in the figure, see the text in
section \ref{sec-atm-vari}}
\label{atmvari}
\end{figure}

We have summarized the variation of atmospheric neutrino fluxes 
for the model variations of primary cosmic ray flux (Fig.~\ref{1ryvari}), 
hadronic interaction(Fig.~\ref{int2vari}),
and atmospheric density structure(Fig.~\ref{atmvari}).

From these figures, one can see that the variation of the primary cosmic 
ray flux and the secondary particle spectrum in the hadronic 
interactions  have a large effect on the atmospheric 
neutrino fluxes. 
However, the variation of other hadronic interaction parameters
and the density structure of atmosphere 
 have far smaller effect on the
atmospheric neutrino fluxes.

\begin{figure}[ht]
\centerline{\epsffile{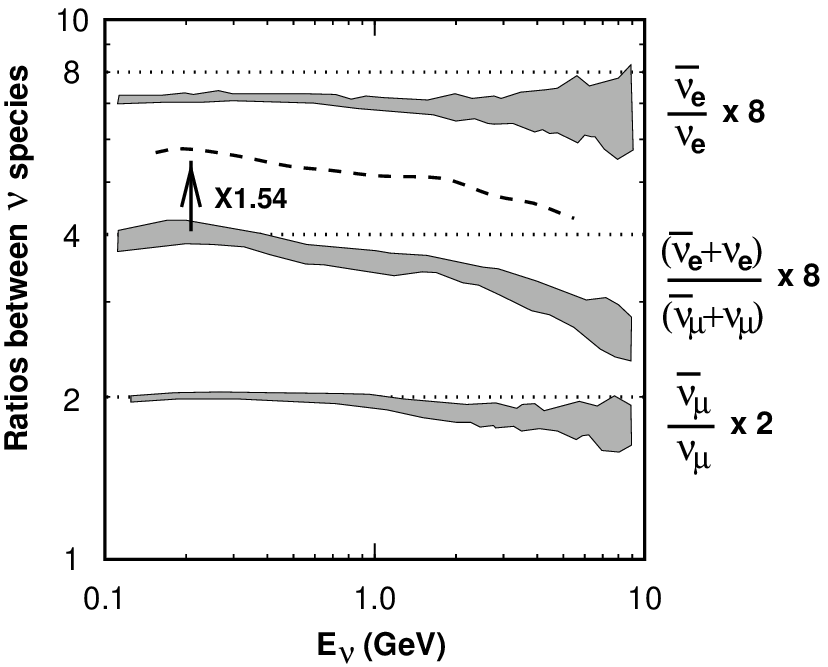}}
\caption{Variation of ratios among different kinds of neutrino
with all the variation considered in sections 
\ref{sec-1ry-vari}--\ref{sec-atm-vari} in all the combinations.
The dashed line shows the ratio 
$(\bar \nu_e + \nu_e)/(\bar \nu_{\mu}+\nu_{\mu})$ 
of HKKM shifted by a factor of 1.54 = ($1/0.65$).
}
\label{ratio-3}
\end{figure}

Shown in Fig.~\ref{ratio-3} is the variation region of the ratios 
among different kinds of neutrinos, with all the variations considered
here. It is seen that the variation of the ratios is very small.
All of the 
$(\bar \nu_e + \nu_e)/(\bar \nu_{\mu}+\nu_{\mu})$
variations are inside of the boundary of $\pm 5$~\% at 1~GeV.
However, the small statistics make the ratio variation larger in 
the $\gtrsim 3$~GeV region due to Monte Carlo method.

We have also depicted
the $(\bar \nu_e + \nu_e)/(\bar \nu_{\mu}+\nu_{\mu})$ ratio by HKKM 
shifted by a factor $1.54 = 1/0.65$.
Note that SuperKamiokande have observed the ratio 
$(\nu_{\mu}/\nu_e)_{obs}/(\nu_{\mu}/\nu_e)_{MC} \approx 0.65$
both for sub-GeV and multi-GeV regions.

\section{Summary and comments}

We have studied the effect of variation in primary cosmic ray flux,
hadronic interaction model, and density structure of atmosphere 
on the atmospheric neutrino fluxes.
The variation of the primary cosmic ray flux is
directly related to the absolute value of atmospheric neutrino fluxes.
The secondary particle spectrum in the hadronic interactions
also strongly affects the absolute value of the fluxes.
Other variations considered here do not have a large effect on 
the atmospheric neutrino fluxes.

Variation of primary cosmic ray flux and/or interaction model does
not cause a large change in the ratio of different kinds of neutrino.
We have noted that the secondary particle spectrum of BN
is outside of our variation limit.
This may be the case for the interaction model used by different authors.
However, the ratio
$(\bar \nu_e + \nu_e)/(\bar \nu_{\mu}+\nu_{\mu})$
does shows a good agreement among different authors
(See the comparison in HKKM).
Probably a variation from a different starting point would not give 
a very different answer.
Thus, it is difficult to explain the ratio 
$(\nu_{\mu}/\nu_e)_{obs}/(\nu_{\mu}/\nu_e)_{MC}$
observed in Kamiokande and SuperKamiokande by the uncertainty 
in the calculation of atmospheric neutrino fluxes.

However, when applying the calculated atmospheric neutrino fluxes to 
the neutrino oscillation study, the absolute values of the fluxes
become important.
The determination of cosmic ray flux to a high accuracy and the 
detailed study of hadronic interaction at 
$x={\rm E}_{2nd}/{\rm E}_{inc}\approx 0.3$ are crucial.

The uncertainty for the arrival direction 
was not discussed so far in this paper.
We shortly comment on the effect of the muon bending by geomagnetic 
field and transverse momentum in the hadronic interaction.
For the muon in the geomagnetic field, we can calculate the average 
bending angle before decay as:
$$
\theta = c \gamma \tau_\mu / r_G 
= { c E_\mu \tau_\mu \over m_\mu} / {E_\mu \over c \cdot eB} 
= c^2 \cdot {eB \over m_\mu} \cdot \tau_\mu
$$
$$
= (3\times 10^{8}({\rm m/sec}))^2 {B({\rm T})\over 106\times 10^6({\rm eV})}
2.2\times 10^{-6}({\rm sec})
$$
$$
 = 0.19 \times 10^4 {B {\rm (T)}}\ . 
$$
Therefore, the bending angle for the typical geomagnetic field 
($\approx 0.5 \times 10^{-4}$~T) is around 0.1 radian or 5 degree.
This is probably not an urgent problem at this moment.

We note that the typical value of the transverse momentum in a hadronic 
interaction is $\sim$ 0.3~GeV.
Therefore, the transverse momentum is unimportant for 
$\gtrsim$ a few GeV.
\vspace{20pt}

\noindent{\bf Acknowledgements.}
The author is grateful to S.~Orito for showing the BESS data in 
Fig.~\ref{le10000}, and J.~Holder for his careful reading of
the manuscript.
He thanks to T.Kajita, K.~Kasahara, and S.~Midorikawa for the 
collaboration, and T.K.~Gaisser for the communications.

\end{document}